\documentclass [12pt]{article}
\usepackage{amsmath}
\usepackage{amssymb}
\usepackage{graphics}
\usepackage{graphicx}

=240mm

\begin{document}
\title{X-Ray Multi-Energy Introscopy Systems with New
Semiconductor Scintillators}
\author{V.D. Ryzhikov, N.G. Starzhinskiy, S.V. Naydenov, \\
E.K. Lisetskaya, L.P. Gal'chinetskii,
and V.I. Silin \\ {\it Institute for Single Crystals of NAS of
Ukraine}, \\ {\it 60 Lenin ave., 61001 Kharkov, Ukraine} }
\maketitle
\begin{abstract}
Theoretical background and data on the ways of practical
realization are presented, related to the problem of detection of
dangerous organic objects (explosives, drugs, etc.) in the
presence of other organic substances with atomic number differing
by no more than 20-30\%. For this purpose, multi-energy X-ray
introscopy is used. It has been shown that the "weakest link" in
the existing multi-energy introscopes used for safety inspection
and medicine are detectors of ionizing radiation.  In particular,
critical is the type of scintillator used in the low-energy
detection subsystem. Data are presented on design principles and
properties of combined detectors based on a new type of
semiconductor scintillators (SCS) -- $ZnSe(Te,O)$, with conversion
efficiency of 19-22\%, afterglow level less then $0.05\, \%$ after
$10\, ms$, and radiation stability up to $500\, Mrad$. Results are
given on the practical use of experimental samples of the
low-energy detector subsystem based on the new SCS material in
two-energy introscopes of the $4$th and $5$th generation.
\end{abstract}

\bigskip

Alarming tendencies towards increased danger of terrorism,
expanding traffic volumes of illegal loads and enclosures of
organic origin (explosives, drugs, components of chemical and
biological weapons, etc.) indicate an increasing need for
improvement of stationary and mobile radiation inspection means
and instrumentation. One of the most promising directions in this
field is development and application of multi-energy
polydimensional (2D and 3D) X-ray introscopy systems (which are
also of great interest for medical express diagnostics).

The existing safety inspection devices (SID) have rather high
ability to detect metal objects and can also reveal organic
materials on the background of inorganic environment. However,
they do not ensure identification of explosives and drugs on the
background of other organic materials, which is an essential
drawback of the existing systems. Contemporary SIDs also generally
fail in such specific applications as detection of organic
inclusions (powders) in postal envelopes.

Possibility of property identification of the imaged objects with
close values of the effective atomic number $Z_{eff}$ by methods
of multi-energy X-ray introscopy has been considered for a
single-channel and N-channel ($N\ge 2$) signal detection methods.
In a standard approach, the structure of an object is synthesized
using ``phantoms'' that correspond to the reference materials with
known properties (e.g., calcium phosphate
$Z_{eff}[Ca_{3}(PO_{4})_{2}]=17.38$ and water
$Z_{eff}[H_{2}O]=7.95$, or carbon $Z_{eff}[C]=6$ and iron
$Z_{eff}[Fe]=26$ -- in two-energy introscopy). Geometrical
dimensions, including absorbing thicknesses $\Delta _{j}$ of the
object elements, are determined by the 2D or 3D configuration of
the detecting system. Reconstruction of the object structure is
made by analysis of the signal amplitude $V_{ i}=\eta
(E_{i})V_{0i}\exp(-\sum _{j}\mu _{ij}\Delta _{j})$, where $\eta $
is ``conversion efficiency'' of the detecting channel, $V _{0i}$
-- power rate of the quasimonochromatic X-ray source with
radiation energy $E_{i}$, $\mu _{ij}$ -- integral (mainly, over
photo- and Compton effects) absorption constant by different
layers $\Delta _{j}$ of the object. Theoretical analysis and
experimental data show that ``synthesis'' of the structure using
single-channel detection can be done only approximately,
qualitatively, not allowing identification of the structure
elements $\Delta _{j}$ with close $Z_{eff}$ values.

More precise analysis of the object structure requires
introduction of additional separate channels for signal detection.
Then even in the simple case of two-energy SID (with 2D- or
3D-configuration of the detecting system) it is possible to
determine $Z_{eff}$ [1] with high precision  for different
elements $\mu _{j}$, and $Z_{eff} = F(V_{ij}; C_{ij}, Z_{j}^{\ast
})$, where $C_{ij}$ and $Z_{j}^{\ast }$ -- are calibration
parameters.

Thus, the key element determining functional parameters of the SID
is a detector array composed of hundreds and thousands of
radiation detectors, which are based on scintillator crystals. The
scintillator characteristics are the main factor that determines
and limits the general sensitivity and detectability of dangerous
inclusions by the detectors and the SID as a whole. Basic
limitations of scintillation parameters (afterglow, light output)
and other properties (radiation hardness and stability,
hygroscopicity, etc.) achieved with the best existing
scintillators -- $CsI(Tl)$, $CWO$, $GSO$, etc. -- are the major
factors that put limits to the capabilities of the existing SID.

Therefore, efforts in engineering and technology should be directed (in
parallel with establishing theoretical base for SID of new generations)
towards development of new types of scintillator crystals.

New prospects in this direction have been commenced by recent
development of a new class of semiconductor scintillators (SCS)
based on $A^{2}B^{6}$ compounds at the Concern ``Institute for
Single Crystals'' (CISC), Kharkov, Ukraine. According to data
X-ray-sensitive elements on the basis of $ZnSe(Te,O)$ have
conversion efficiency of 0.19-0.22 exceeding that of $CsI(Tl)$
(0.15), technical light output of up to 120-150\% with respect to
$CsI(Tl)$, their afterglow level after $10-20\,ms$ is by 1-2
orders of magnitude lower ($ \le 0.03\% $) and radiation stability
is more than 1000 times higher than those of $CsI(Tl)$ [2].
Preliminary data show that conductivity of scintillators of this
type can be intentionally varied within 10 to 14 orders of
magnitude. This might be important for development of
scintillation detectors with surface-integrated photosensitive
heterostructures based on $A^{2}B^{6}$ compounds [3].

To construct a two-energy X-ray introscopic SID, we chose
$ZnSe(Te,O)$ crystals as scintillation elements for the low-energy
detector subsystem of ``scintillator
$ZnSe(Te,O)$--$Si$-photodiode'' type. Among all known
scintillators used in X-ray introscopy ($CsI(Tl)$, $CWO$, $GSO$,
etc.) SCS $ZnSe(Te,O)$ have the best complex of parameters, which
ensure for detectors based on these crystals the highest absolute
radiation sensitivity (especially in the low energy range of
$E_{i}< 50-70 keV$), steepness of $dV/d\rho $ signal
transformation, as well as the broadest dynamic range due to high
conversion efficiency of this SCS and low afterglow level.

It is also interesting to note that relatively low value of $Z_{
eff} [ZnSe(Te,O)] \approx 32$, which is generally considered as
disadvantage for a scintillator, is an important positive factor
from the viewpoint of multienergy X-ray spectroscopy. This
property allows SCS $ZnSe(Te,O)$ to be used in ``sequentially
located'' two-energy detectors simultaneously in two different
roles -- as a scintillation element and as a low-energy radiation
filter with low values of accumulation factor and scattering
constants. In the high-energy detector subsystem, depending upon
SID purpose, conventional scintillators -- $CsI(Tl)$, $CWO$, etc.
can be used.

Data are presented on the practical use of experimental samples of
the low-energy detector subsystem based on the new SCS material in
two-energy introscopes of the $4^{th}$ and $5^{th}$ generation by
leading companies in Germany and Ukraine. Their detecting ability
towards potentially dangerous organic inclusions has been shown to
be substantially superior, with penetrating ability (steel)
150-200\% higher, as compared with similar introscopes produced by
leading USA companies, which used other kinds of scintillators.

\bigskip

[1]~S.V.~Naydenov, V.D.~Ryzhikov, ``Determining Chemical
Compositions by Method of Multi-energy Radiography''
\textit{Technical Physics Letters}, vol.~28, 2002, pp.~357-360.

[2]~V.D.~Ryzhikov, N.G.~Starzhinskiy, L.P.~Gal'chinetskii,
M.~Guttormsen, A.A.~Kist, and V.~Klamra, ``Behavior of New
ZnSe(Te,O) Semiconductor Scintillators Under Doses of Ionizing
Radiation'', \textit{IEEE Trans. Nucl. Sci.,} vol.~48, \# 4, 2001,
pp.1561-1564.

[3]~A.I. Focsha, P.A.~Gashin, V.D.~Ryzhikov, and
N.G.~Starzhinskiy, ``Preparation and Properties of an Integrated
System ``Photosensitive Heterostructure-Semiconductor
Scintillator'' on the Basis of Compounds A$^{{\rm I}{\rm
I}}$B$^{{\rm V}{\rm I}}$, \textit{Intern. J. Inorg. Mater.,} \# 3,
2001, pp. 1223-1226.

\end{document}